\documentclass[sigconf, nonacm]{acmart}
\usepackage{enumerate}
\usepackage{multirow}
\usepackage{balance}

\newcommand\vldbyear{2024}
\newcommand\vldbworkshop{Cloud Databases (CloudDB)}
\newcommand\vldbauthors{\authors}
\newcommand\vldbtitle{\shorttitle} 
\newcommand\vldbavailabilityurl{}
\newcommand\vldbpagestyle{empty} 

\begin{document}
\title{MetaHive: A Cache-Optimized Metadata Management for Heterogeneous Key-Value Stores}

%%
%% The "author" command and its associated commands are used to define the authors and their affiliations.
\author{Alireza Heidari}
\email{alireza.heidarikhazaei@huawei.com}
\affiliation{%
  \institution{Huawei Cloud}
  % \streetaddress{}
  % \city{Vancouver}
  % \state{BC}
  % \country{Canada}
  % \postcode{43017-6221}
}

\author{Amirhossein Ahmadi}
\email{amirhossein.ahmadi@huawei.com}
\affiliation{%
  \institution{Huawei Cloud}
  % \streetaddress{}
  % \city{Vancouver}
  % \state{BC}
  % \country{Canada}
  % \postcode{43017-6221}
}

\author{Zefeng Zhi}
\email{zefeng.zhi@huawei.com}
\affiliation{%
  \institution{Huawei Cloud}
  % \streetaddress{}
  % \city{Vancouver}
  % \state{BC}
  % \country{Canada}
  % \postcode{43017-6221}
}

\author{Wei Zhang}
\email{wei.zhang6@huawei.com}
\affiliation{%
  \institution{Huawei Cloud}
  % \streetaddress{}
  % \city{Vancouver}
  % \state{BC}
  % \country{Canada}
  % \postcode{43017-6221}
}

%%
%% The abstract is a short summary of the work to be presented in the
%% article.
\begin{abstract}
Cloud key-value (KV) stores provide businesses with a cost-effective and adaptive alternative to traditional on-premise data management solutions. KV stores frequently consist of heterogeneous clusters, characterized by varying hardware specifications of the deployment nodes, with each node potentially running a distinct version of the KV store software. This heterogeneity is accompanied by the diverse metadata that they need to manage. 
In this study, we introduce MetaHive, a cache-optimized approach to managing metadata in heterogeneous KV store clusters.
MetaHive disaggregates the original data from its associated metadata to promote independence between them, while maintaining their interconnection during usage. This makes the metadata opaque from the downstream processes and the other KV stores in the cluster.
MetaHive also ensures that the KV and metadata entries are stored in the vicinity of each other in memory and storage. This allows MetaHive to optimally utilize the caching mechanism without extra storage read overhead for metadata retrieval. We deploy MetaHive to ensure data integrity in RocksDB and demonstrate its rapid data validation with minimal effect on performance.

\end{abstract}

\maketitle

%%% do not modify the following VLDB block %%
%%% VLDB block start %%%
\pagestyle{\vldbpagestyle}
\begingroup\small\noindent\raggedright\textbf{VLDB Workshop Reference Format:}\\
\vldbauthors. \vldbtitle. VLDB \vldbyear\ Workshop: \vldbworkshop.\\ %\vldbvolume(\vldbissue): \vldbpages, \vldbyear.\\
%\href{https://doi.org/\vldbdoi}{doi:\vldbdoi}
\endgroup
\begingroup
\renewcommand\thefootnote{}\footnote{\noindent
This work is licensed under the Creative Commons BY-NC-ND 4.0 International License. Visit \url{https://creativecommons.org/licenses/by-nc-nd/4.0/} to view a copy of this license. For any use beyond those covered by this license, obtain permission by emailing \href{mailto:info@vldb.org}{info@vldb.org}. Copyright is held by the owner/author(s). Publication rights licensed to the VLDB Endowment. \\
\raggedright Proceedings of the VLDB Endowment. %, Vol. \vldbvolume, No. \vldbissue\ %
ISSN 2150-8097. \\
%\href{https://doi.org/\vldbdoi}{doi:\vldbdoi} \\
}\addtocounter{footnote}{-1}\endgroup
%%% VLDB block end %%%

%%% do not modify the following VLDB block %%
%%% VLDB block start %%%
\ifdefempty{\vldbavailabilityurl}{}{
\vspace{.3cm}
\begingroup\small\noindent\raggedright\textbf{VLDB Workshop Artifact Availability:}\\
The source code, data, and/or other artifacts have been made available at \url{\vldbavailabilityurl}.
\endgroup
}
%%% VLDB block end %%%

\section{Introduction}
\label{sec:intro}
\noindent 
\textbf{Context.} The global cloud computing market was valued at \textit{USD 602.31 billion} in 2023 and is projected to grow at a \textit{rate of 24.8\%} from 2024 to 2030~\cite{cloudmarket}. Cloud computing provides businesses with a cost-effective and adaptive alternative to traditional on-premise data management solutions. Key-value (KV) stores represent a substantial portion of the market, compromising \textit{over 20\%} of the Cloud Databases sector~\cite{nosqlmarket}.

A KV store is a form of non-relational database that stores data in pairs of key values, each key acting as a unique identifier and projecting it to the corresponding value~\cite{FUSEE}. The flexibility of this model allows for the storage of a diverse range of data, from simple objects to intricate compound objects. One of the primary benefits of KV stores is their high partitionability, which allows horizontal scaling that exceeds the capabilities of other database models~\cite{zhou2023foundationdb}. In the past decade, \textit{over 20 KV stores} have been developed, including LevelDB~\cite{leveldb}, RocksDB~\cite{dong2021rocksdb}, 
 and DuckDB~\cite{Duckdb}. These databases are widely used by cloud providers, handling billions of data points across various cloud platforms. \footnote{In general, MetaHive works on all databases that store keys in a sorted manner: LevelDB~\cite{leveldb}, RocksDB~\cite{dong2021rocksdb}, WiredTiger (used by MongoDB)~\cite{membrey2010definitive}, BadgerDB, TiKV, and DuckDB~\cite{Duckdb}.
}

Key-value metadata~\cite{FUSEE} is essential in various situations in cloud databases, where it requires storing and utilizing additional information related to the specified key and value in downstream application processes. An example is the \textit{ETL (Extract, Transform, and Load) process}~\cite{etl}, where data from various sources are combined into a centralized repository known as a data warehouse. Metadata is heavily relied upon in this process~\cite{etlmetadata}, as it facilitates the application of business rules to clean, organize, and prepare raw data for storage, data analysis, and machine learning (ML) applications.
Another significant example in the Cloud DB context is \textit{verifying data integrity}. Cloud KV stores often form a distributed KV cluster due to the substantial volume of data and processing involved~\cite{yang2021large}. As these clusters evolve, they encompass a diverse range of nodes with varying hardware capabilities, each running a distinct version of the database software code, resulting in heterogeneous KV clusters. In such environments, verifying the accuracy of the data is essential, as the data frequently migrate between various nodes, and the hardware or software characteristics of each node may introduce errors in the KV store entries. 

\noindent\textbf{Motivational Example.} Consider a cluster of three nodes $N_1$, $N_2$, and $N_3$ within a banking system, interconnected via communication protocols and APIs. However, these nodes are not identical; the cluster is heterogeneous where $N_1$ and $N_3$ share the same setup, while $N_2$ is an older setup with more error-prone devices. In this scenario, the cloud owner opts to prevent error injection by implementing a data integrity system (e.g., checksum) on $N_2$ such that $N_1$ and $N_3$ remain unaware of its presence. Due to the ease of correcting minor errors, $N_2$ necessitates a high-performance data integrity system operating at the finest granularity. Additionally, there should be no changes in the API or the code of the other nodes. Given that the data stored in this cloud comprises high-profile banking information, the integrity system must not alter the values' nature. MetaHive offers a management solution that allows the cloud owner to satisfy all these requirements concurrently.

\noindent\textbf{Objectives.} Three main objectives should be considered for metadata management in cloud KV stores:
\begin{enumerate}[(i)]
    \item \textbf{Performance}: The metadata for each KV should be located close to its corresponding data because, in many instances, the application requires access to the metadata immediately after reading the KV (e.g., verifying data correctness). To enhance cache efficiency and minimize cache misses and memory page (block) lookups, the metadata should be placed on the same memory block as the corresponding KV.
    \item \textbf{Heterogeneity}: Cloud databases often employ a distributed KV store architecture, where each KV store shard is hosted on a separate node. These nodes can have varying hardware specifications, and the software version of the KV store on each node might differ. Within this diverse KV store cluster, it is crucial to ensure that introducing the KV metadata does not affect any version of KV store applications. This necessitates \textit{backward compatibility}, allowing older KV data to function with current data, as well as \textit{forward compatibility}, ensuring that the new version of KV operates with previous software code.
    \item \textbf{Privacy}: In a clustered environment, each KV shard on a node contains a subset of all KVs. These KVs could represent private data specific to each node. Consequently, the metadata containing information about the key values of that node should be stored on the same edge node and should not be migrated to other shards. 
\end{enumerate}

\noindent\textbf{Contributions.} In this paper, we present \textit{MetaHive, a metadata management solution designed with privacy and efficiency in mind for diverse KV stores}, which uniquely addresses all these objectives at once. While the proposed design is applicable to all KV stores, it is practically implemented and evaluated on a heterogeneous cluster of RocksDB nodes, a high-performance embedded database for key-value data. The focus is on ensuring data integrity in a cluster of RocksDB shards that constitute a Cloud KV store, where the DB provider must assure users of data correctness~\cite{ghazizadeh2013data}.

MetaHive introduces metadata as a KV structure that is specifically designed to work in a heterogeneous cluster while maintaining backward and forward compatibility. The design also ensures the privacy of each node's metadata. To ensure data integrity, a checksum part is included in the payload of the metadata entries. This checksum represents the checksum of the corresponding KV entry, providing a means to verify the integrity of the data. Additionally, MetaHive guarantees that both the KV store pairs and their corresponding metadata are written on the same memory page throughout various RocksDB processes, such as compaction. This ensures efficient and consistent handling of data and metadata within the system.

\noindent\textbf{Organization.} The remainder of the paper proceeds as follows: Section \ref{sec:background} covers the background concepts. Section \ref{sec:relatedWork} discusses related work, and Section \ref{sec:overview} presents an overview of our proposed architecture. In Section \ref{sec:eval}, we evaluate the implementation of our solution in RocksDB, and Section \ref{sec:conclusion} concludes with a summary of the key points.

\section{Background}
This section provides background information for the study and outlines the essential terminology and concepts used in this work.
\label{sec:background}
\subsection{Block, Memory Page, and Cache Optimality}
\label{sec:background:cacheOptimality}
A memory block is a contiguous chunk of memory allocated for specific purposes in a computer system, managed by the operating system, and used to store data or instructions. Block size, a critical parameter in file systems and databases, affects data access and storage efficiency. A memory page is a contiguous block of virtual memory described by a single entry in the page table, typically sized in powers of two (e.g. $4$ or $8$ KB), and is crucial for memory management. Cache optimality refers to the degree to which an algorithm or data structure uses cache memory to frequently access data within the same memory page to minimize page faults and improve performance\cite{mem1,mem2,yapar2019optimality}. Designing with memory blocks and pages in mind can improve cache usage and system efficiency.

\begin{figure}[t]
  \centering
  \makebox[\columnwidth][c]{\includegraphics[width=\columnwidth]{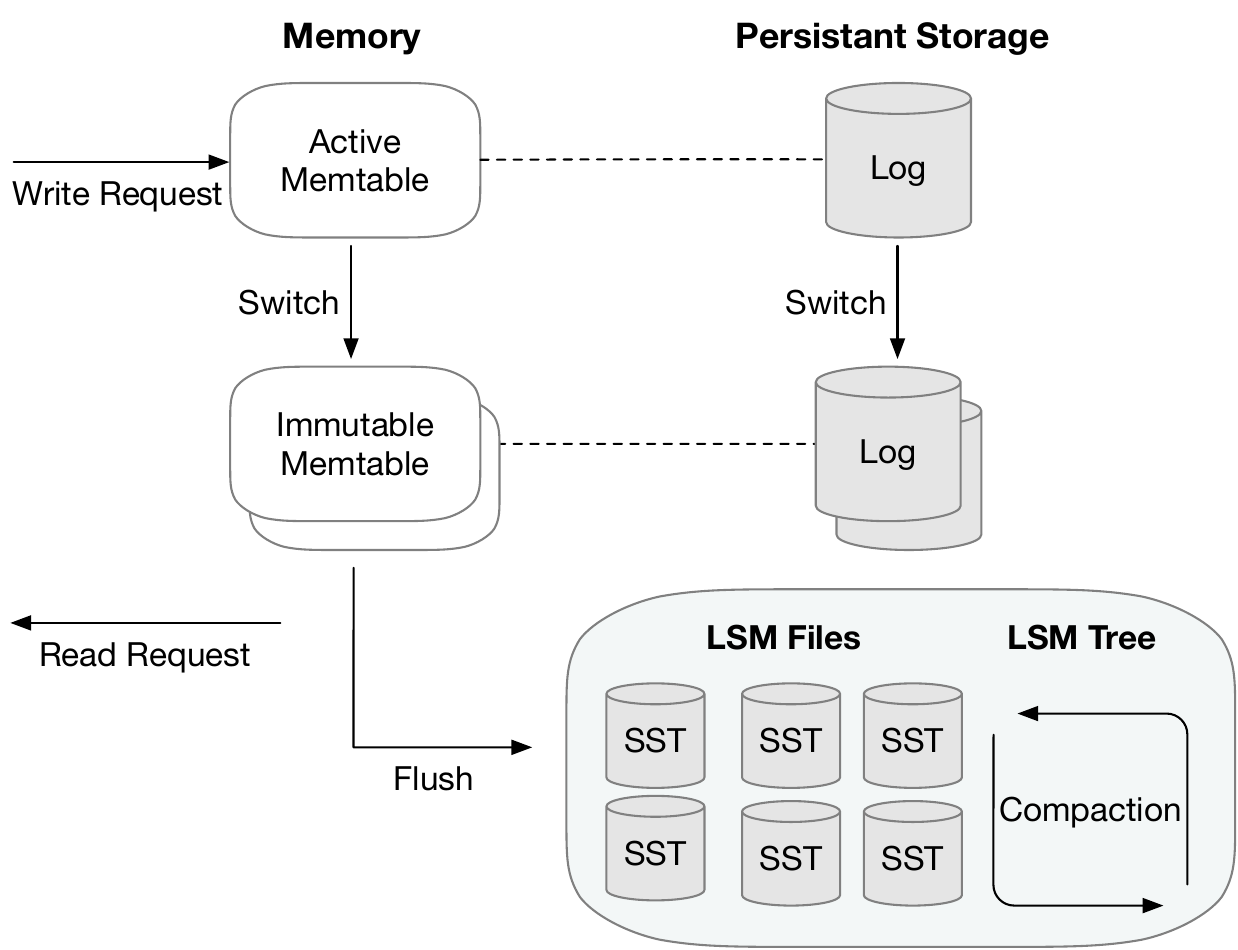}}
  \caption{RocksDB Architecture}
  \label{fig:RocksDB-arch}
\end{figure}

\subsection{RocksDB Architecture}
\label{sec:rocksdb}
RocksDB, created by Meta in 2012, is a highly efficient key-value storage engine~\cite{rocksdbpaper}. It is specifically optimized to utilize the capabilities of Solid State Drives (SSDs). This storage engine is mainly designed for large-scale distributed applications and is implemented as a library component that can be integrated into higher-level applications. RocksDB uses log-structured merged trees (LSM) as its core data structure \cite{LSm}.

When data is written to RocksDB, it undergoes two main processes. Initially, the data are placed into an in-memory write buffer known as the MemTable. At the same time, a Write-Ahead Log (WAL) is generated on the disk (see Figure \ref{fig:RocksDB-arch}). The MemTable is structured as a skiplist, maintaining the data in an ordered fashion with an insertion and search complexity of $O(log n)$. The WAL acts as a recovery tool in the event of failures, although its use is optional. Once the MemTable hits a predetermined size threshold, both it and the WAL are set to an immutable state. Then, new MemTable and WAL instances are created for future writes. The data in the MemTable are then transferred to a 'Sorted String Table' (SST) file on the disk, and the old MemTable and WAL are discarded. Each SST organizes data in a sorted sequence and is divided into equally sized blocks. Furthermore, each SST contains an index block that contains one index entry per data block, enabling efficient binary search operations.

The LSM tree in RocksDB is made up of several levels, each of which is built from multiple SSTs. The newest SSTs are generated by flushing MemTables and placed in Level-0. Levels higher than Level-0 are generated through a process called compaction~\cite{compaction}. The size of the SSTs on each level is restricted by configurable parameters. If the target size for a specific level, such as Level-L, is exceeded, a subset of Level-L SSTs is chosen and combined with overlapping Level-(L + 1) SSTs. This compaction process eliminates deleted and overwritten data, optimizing the table for improved read performance and space efficiency. Consequently, the written data are gradually migrated from Level-0 to the highest level. The compaction I/O operations are efficient as they can be parallelized and involve bulk reads and writes of entire files. However, this process requires fetching multiple data from the disk and creating new SSTs, which are then written back to the disk. These operations are prone to errors caused by software and hardware failures.

\vspace{-0.5em}
\subsection{Heterogeneous KV store Cluster}
\label{sec:hetro}
As the operational load of the system grows, vertically scaling a single node might not be enough to solve scalability problems. This necessitates the formation of a cluster of KV stores. Within this cluster, various KV store applications are installed on distinct nodes, allowing them to manage the increased loads together. These nodes interact with one another or with a dispatcher edge node to efficiently address queries. Privacy concerns also motivate the use of KV store clusters. When various applications or users need access to certain parts of the data or specific KV nodes, clustering offers essential isolation and management. Dividing the data among the clusters allows each application or user to reach their specific portion of the data, guaranteeing privacy and data separation.

KV store clusters frequently consist of heterogeneous clusters, characterized by varying hardware specifications of the deployment nodes, with each node potentially running a distinct version of the KV store software. The diversity in hardware and software versions complicates the process of updating the KV store. Updates to the KV store software must be compatible with the data of the current versions running on other nodes in the cluster. Incompatibility could lead to cluster failures, resulting in interruptions in data access and availability. Thus, when adding metadata per key-value in a KV store cluster, it is crucial to design the system so the data remains unobservable from other shards or nodes. This ensures both privacy and compatibility, facilitating efficient data management and access within the cluster while preserving the system's overall integrity and consistency.

\section{Existing Data Integrity Methods for KV stores}
\label{sec:relatedWork}

In this section, we focus on maintaining the integrity of key-value data, using it as an example to demonstrate the integration of metadata for each KV store pair. To elucidate these solutions and make them more concrete, we illustrate them using RocksDB. Currently, RocksDB employs two main methods to verify the integrity of the data: (1) incorporating a checksum into the KV payload and (2) embedding a checksum within the metadata block. However, these methods face challenges related to heterogeneous clusters, privacy issues, and performance enhancement.

\begin{figure}[t]
  \centering
  \makebox{\includegraphics[width=0.7\columnwidth]{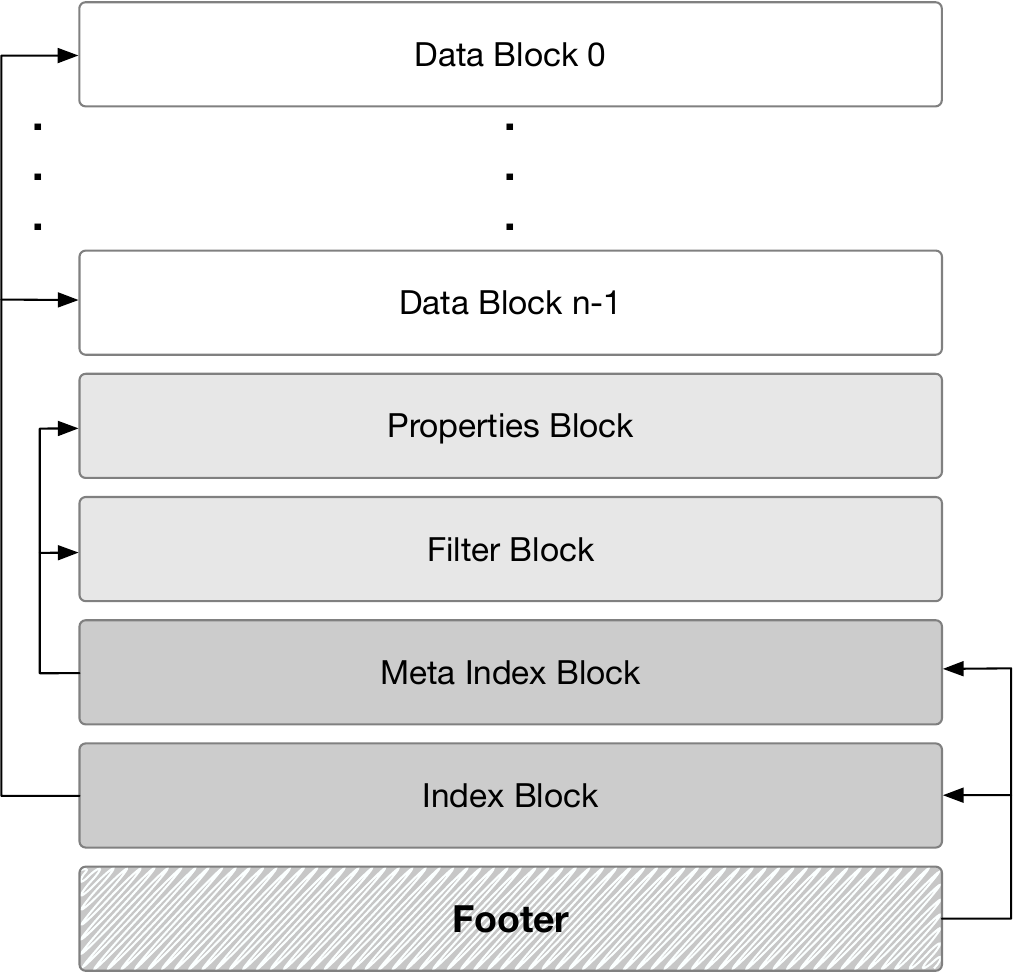}}
  \caption{RocksDB block-based SST format}
  \label{fig:sst-format}
\end{figure}

\subsection{Adding checksum to the KV payload}
\label{sec:exs-payload}
A typical technique involves adding a checksum as metadata to the end of each KV value payload. Upon querying for data, the KV store strips the checksum and delivers the value to the user. Although this method appears straightforward and efficient for handling any metadata, it has some drawbacks.

First, it is incompatible with heterogeneous RocksDB clusters. This issue stems from changes in the data structure that other versions of RocksDB cannot interpret. If SST files are migrated from a RocksDB version with the KV checksum to an older version that does not recognize it, the older version will misinterpret the checksum as part of the value payload, leading to incorrect data interpretation. Furthermore, consider other types of key-value metadata, such as the statistics needed for ETL \cite{statistics}, such as the KV distance from the median. In these cases, retrieving the metadata value requires loading the entire value payload from storage to memory and processing it. RocksDB payloads can be as large as 3GB, but processing the actual value for these metadata types is unnecessary, causing avoidable overhead. Furthermore, this process alters the user value and requires enough information to determine the end of the value, which could pose security risks. Additionally, this method requires shared information on the extraction of value and metadata among all cluster nodes, which contradicts the principle of heterogeneity.

\subsection{Adding checksum to the Footer Block}
An alternative method to incorporate the checksum metadata while preserving the value payload is to append the checksum to the SST footer. Figure~\ref{fig:sst-format} illustrates a block-based SST structure. It comprises multiple data blocks containing key-value pairs, which are separated to enhance cache optimization for KV retrieval. The final block of each SST is a footer that can be extended to include various types of data. To ensure data integrity, the computed metadata (e.g., checksum) can be added to the footer block. Consequently, the key-value pairs remain unchanged, making this approach suitable for heterogeneous RocksDB clusters.

The issue with placing a checksum in the footer block is that it is not optimized for caching, as it disaggregates the metadata from the target KVs on different memory pages (see Section \ref{sec:background:cacheOptimality}). Consequently, whenever we need to obtain KV metadata, such as checking its checksum, we have to access both the footer block and the data block, resulting in two memory page accesses. As the SST size increases, this problem is exacerbated, increasing processing costs. This method incurs a significant memory read overhead, as it requires loading two pages into memory for each KV.

\section{MetaHive Design}
\label{sec:overview}

In this section, we present MetaHive~\footnote{A data integrity mechanism created following the MetaHive framework, with a patent application filed (U.S. Patent Serial No. 18/769665)}, \textit{a cache-optimized method designed to store metadata in KV stores}, considering a heterogeneous KV store cluster. Although MetaHive is a versatile approach for managing metadata, we specifically apply it for data integrity and error detection of RocksDB data. We define error detection as identifying the discrepancies between the stored value and the original value \cite{heidari2019holodetect}, which involves storing checksum metadata per key-value. 

\begin{figure}[t]
  \centering
  \makebox[\columnwidth][c]{\includegraphics[width=\columnwidth]{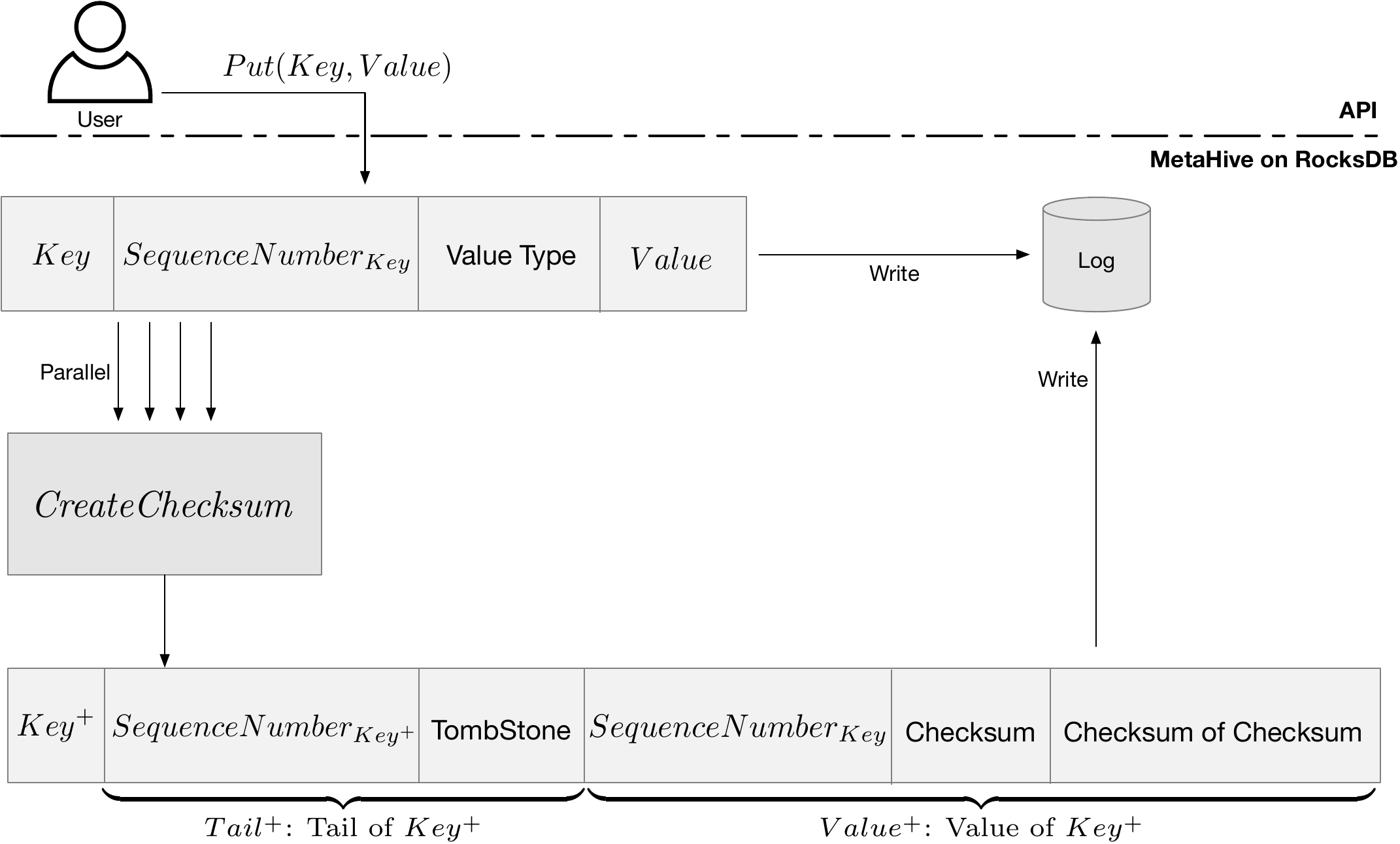}}
  \caption{Inserting checksum metadata on PUT operation}
  \label{fig:put-scenario}
\end{figure}

\subsection{Creating Checksum Metadata}
We need to describe the key-value entry of RocksDB before explaining how we design MetaHive \textit{to disaggregate metadata from their corresponding data}\cite{edara2021big,carvalho2023state,jin2024efficient}. RocksDB appends a 56-bit integer \textit{sequence number} to know the order of the entries with similar keys, and an 8-bit integer indicating the specific \textit{operation type} of the entry (e.g., PUT, Delete, Merge). MetaHive uses the RocksDB entry design and creates metadata (Figure~\ref{fig:put-scenario}) as separate KVs to be stored close to the original data. We calculate the checksum of the corresponding KV and add it to the metadata payload to verify data integrity.

\subsubsection{Metadata Key Generation}
\label{sec:key-cluster}
To ensure that the associated metadata is placed immediately after its KV in both the MemTable and the underlying SST files, we append a special character called the "Start of Heading" (\texttt{SOH}) symbol, represented by `$\backslash001$', to the end of the original key. This special character acts as the metadata identifier and we impose the rule that standard keys do not terminate with this character. Additionally, it is generally known that many APIs and systems already reject such keys. Since the \texttt{SOH} symbol is the smallest character following the \texttt{NUL} symbol (`$\backslash000$'), which is also prohibited at the end of strings on all systems, it ensures that the keys and their associated checksum metadata are placed consecutively. As illustrated in Figure \ref{fig:put-scenario}, the metadata KV record's key is represented as $Key^+$, and is followed by a $Tail^+$ that includes the typical sequence number and value type for the metadata KV record.

\subsubsection{Metadata Payload Generation}
We calculate the hash values of the key, value, sequence number, and type individually. These values are then XORed together and added to the checksum payload. In addition, we compute the checksum of the checksum to verify the integrity of the checksum itself at a later stage. The final payload contains the sequence number of the original key, the checksum value, and the checksum of the checksum. We use XXhash3 as the hash function, which is a fast hash algorithm with processing at RAM speed limits. All these variables constitute the value component of the key-value pair, which includes the checksum information depicted in Figure \ref{fig:put-scenario} as $Value^+$. 

\subsection{KV and Metadata Clusters}
\begin{figure}[t]
  \centering
  \makebox[\columnwidth][c]{\includegraphics[width=\columnwidth]{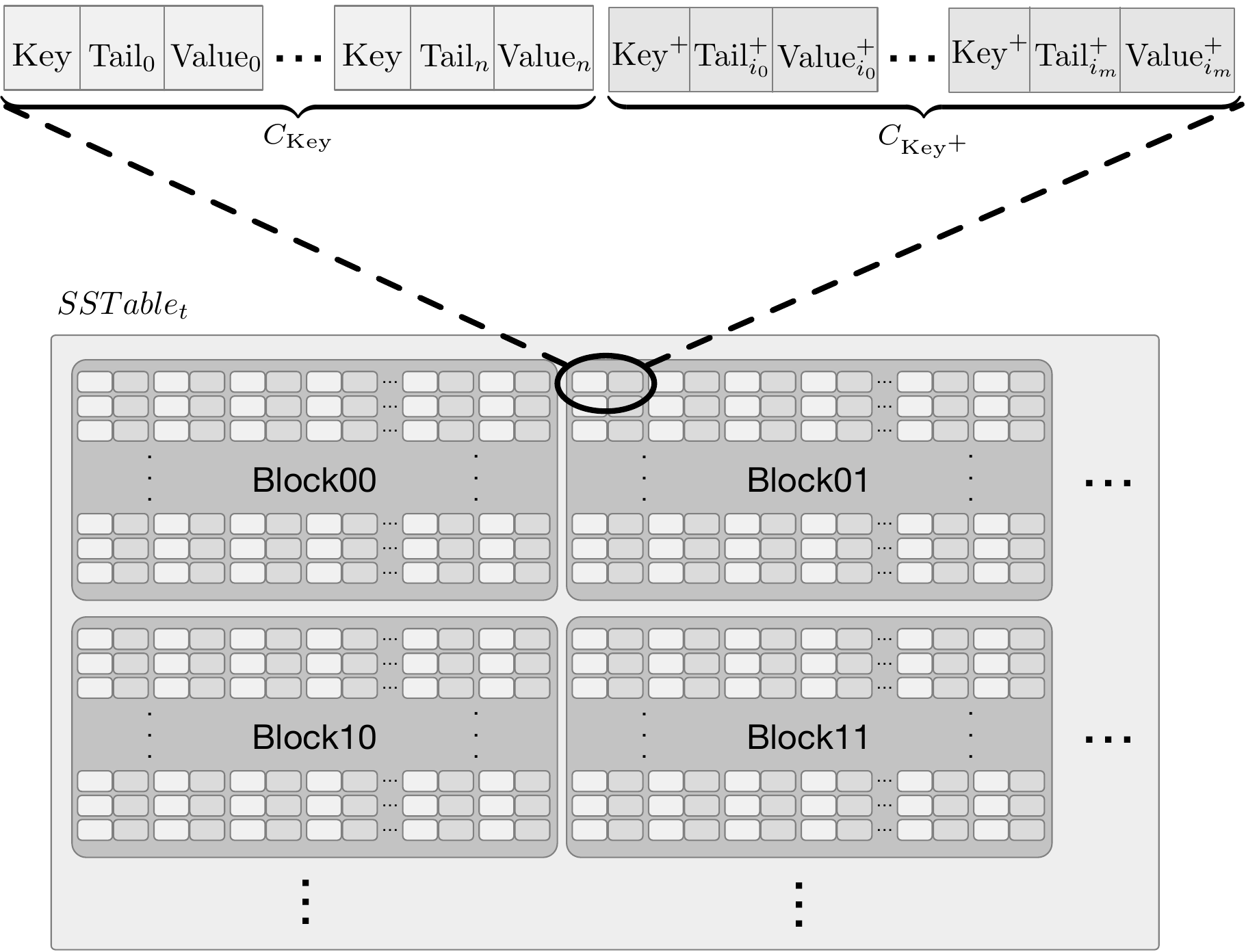}}
  \caption{Clusters of KV and corresponding metadata}
  \label{fig:key-cluster}
\end{figure}

The keys and their corresponding checksums can generate \textit{two sequential clusters} as shown in Figure~\ref{fig:key-cluster}. When multiple updates are applied to the same key and these updates are stored in the same MemTable or SST file, a cluster of identical keys $C_{Key}$ will be formed. When the checksum metadata for a KV pair, $K^+V$, is added to the MemTable for all KV records, the checksum pairs that share the duplicate key $Key^+$ form a cluster $C_{Key^+}$ immediately following the cluster $C_{Key}$. In such cases, it is necessary to determine which checksum key corresponds to each original key. This is achieved by including the sequence number of the original key, which is a unique number, in the initial part of the checksum payload $Value^+$ (see Figure \ref{fig:put-scenario}).

\subsubsection{Cluster Analysis} The sequence of keys in $C_{Key}$ does not correspond directly to the sequence of keys in $C_{Key^+}$. This is because some keys are deleted or filtered, but their metadata is in the SST, which we call \textit{orphan metadata}. There might also be some keys that do not have any corresponding metadata (originally created from nodes without the MetaHive design). MetaHive uses a single-pass algorithm to correspond the KVs with their metadata (Section~\ref{sec:alg}). MetaHive also removes orphan metadata to enhance storage efficiency, since the associated keys for these metadata have already been deleted.

\subsubsection{Cache-Optimized Metadata Retrieval}
MetaHive put the metadata entry in the vicinity of the actual entries by appending the specific character (Section~\ref{sec:key-cluster}). However, this does not guarantee that entries and their checksum are written in the same SST block to be efficiently retrieved in memory by reading one SST block file. RocksDB determines the maximum size of the block as a constant and dumps entries into the block files in the compaction process (see Section~\ref{sec:rocksdb}) whenever it reaches this capacity. This process may separate some entries with their corresponding metadata. We modify the RocksDB code to ensure that the actual KV entries and their corresponding checksum flush in the same block file during the compaction process. Therefore, reaching the metadata of an entry does not require reading two block files.

\subsection{Metadata within a Heterogeneous Cluster}
The significant goal of MetaHive is to achieve data integrity in a heterogeneous cluster (Section~\ref{sec:hetro}). In our scenario, each node might run a different version of the RocksDB code. Hence, it is vital to ensure that nodes without checksum metadata support are unable to detect it. It is important that the checksum metadata stays hidden, guaranteeing that (i) key iteration skips the checksum metadata, and (ii) automatic cleanup happens when SST files from the MetaHive version of RocksDB are transferred to an older version of RocksDB.

\noindent\textit{Tombstone to Skip and Auto-cleaning Metadata} LSM-based KV stores, such as RocksDB, frequently employ a particular Tombstone type for key deletion. This type ensures that the keys marked with it are automatically purged from the lower SST levels, and their iterator bypasses these keys. Hence, using Tombstone, we fulfill the two objectives in the diverse cluster on the RocksDB nodes that do not include the MetaHive design. It is important to note that we cannot introduce a new type to RocksDB because it would not be recognized by older versions of RocksDB. MetaHive uses the following criteria to identify a checksum metadata entry:
$$
is\_metadata = key.endsWith(``\backslash001") \wedge key.type == Tombstone
$$

\subsection{Data Integrity Check during Compaction}
\label{sec:alg}
RocksDB is a write-intensive application designed for rapid data ingestion. To ensure consistency and fuse \cite{fusion} duplicated data, it periodically performs a compaction process, selecting data from both memory and disk to create a new sorted structure. This process is prone to errors, as numerous KV reads and writes from memory and storage occur during this process. In this process, we first check the integrity of the data using the checksum metadata and then we use a repair mechanism for erroneous data.

\begin{figure}[t]
  \centering
  \makebox[\columnwidth][c]{\includegraphics[width=\columnwidth]{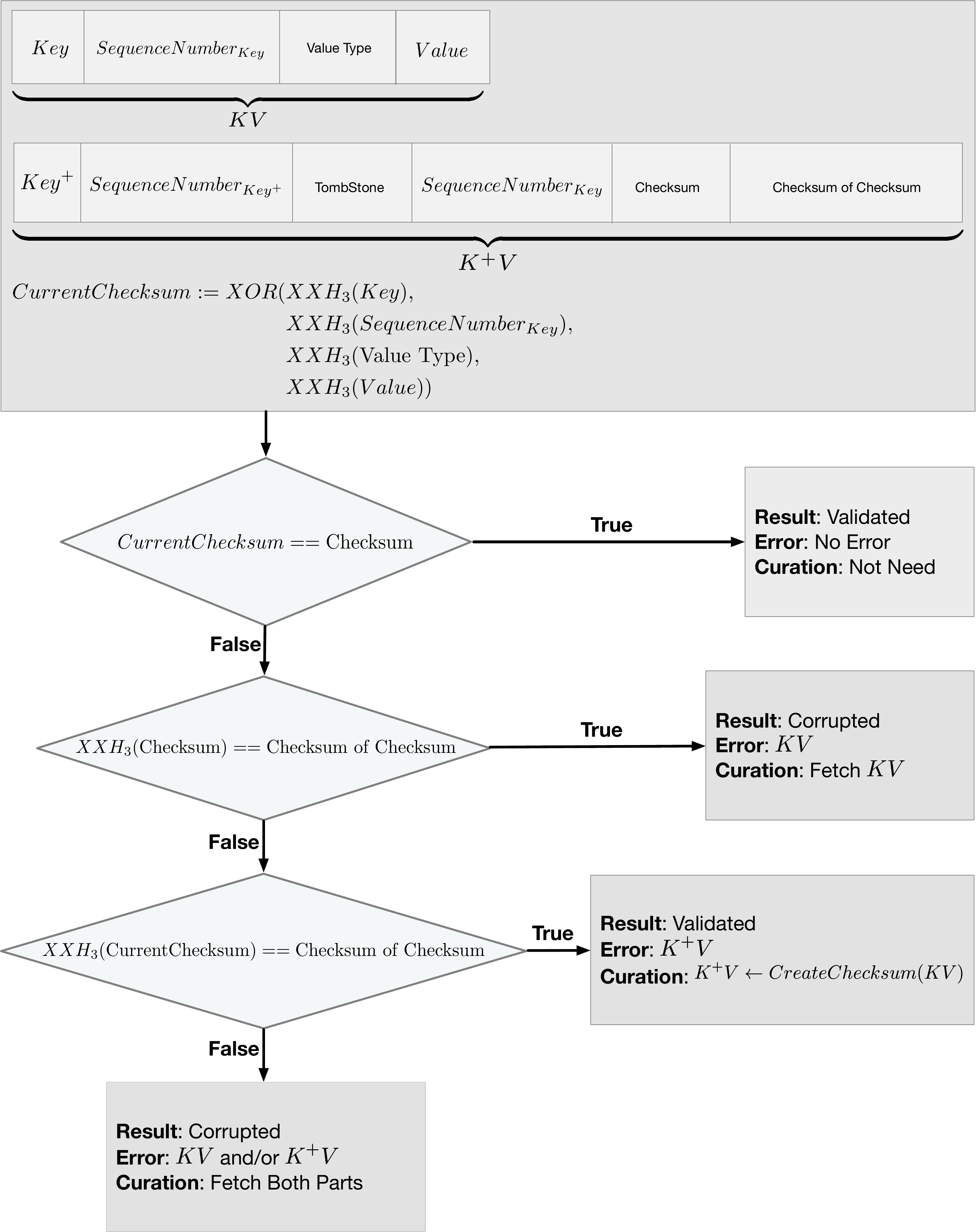}}
  \caption{$DataIntegrity(KV,K^+V)$ the MetaHive error detection module for data integrity procedure.}
  \label{fig:compaction}
\end{figure}

\subsubsection{Error Detection Module} We utilize a single-pass approach for error detection. For each key within the cluster $C_{Key}$ and its associated metadata cluster $C_{Key^+}$, we first traverse $C_{Key}$, compute their checksums, and store them in a temporary map data structure using the key sequence number of the KV. Subsequently, we use the sequence number stored in $Value^+$ of each $K^+V$ in $C_{Key^+}$ to retrieve the corresponding key from the map.

\begin{figure}[t]
  \centering
  \makebox[\columnwidth][c]{\includegraphics[width=\columnwidth]{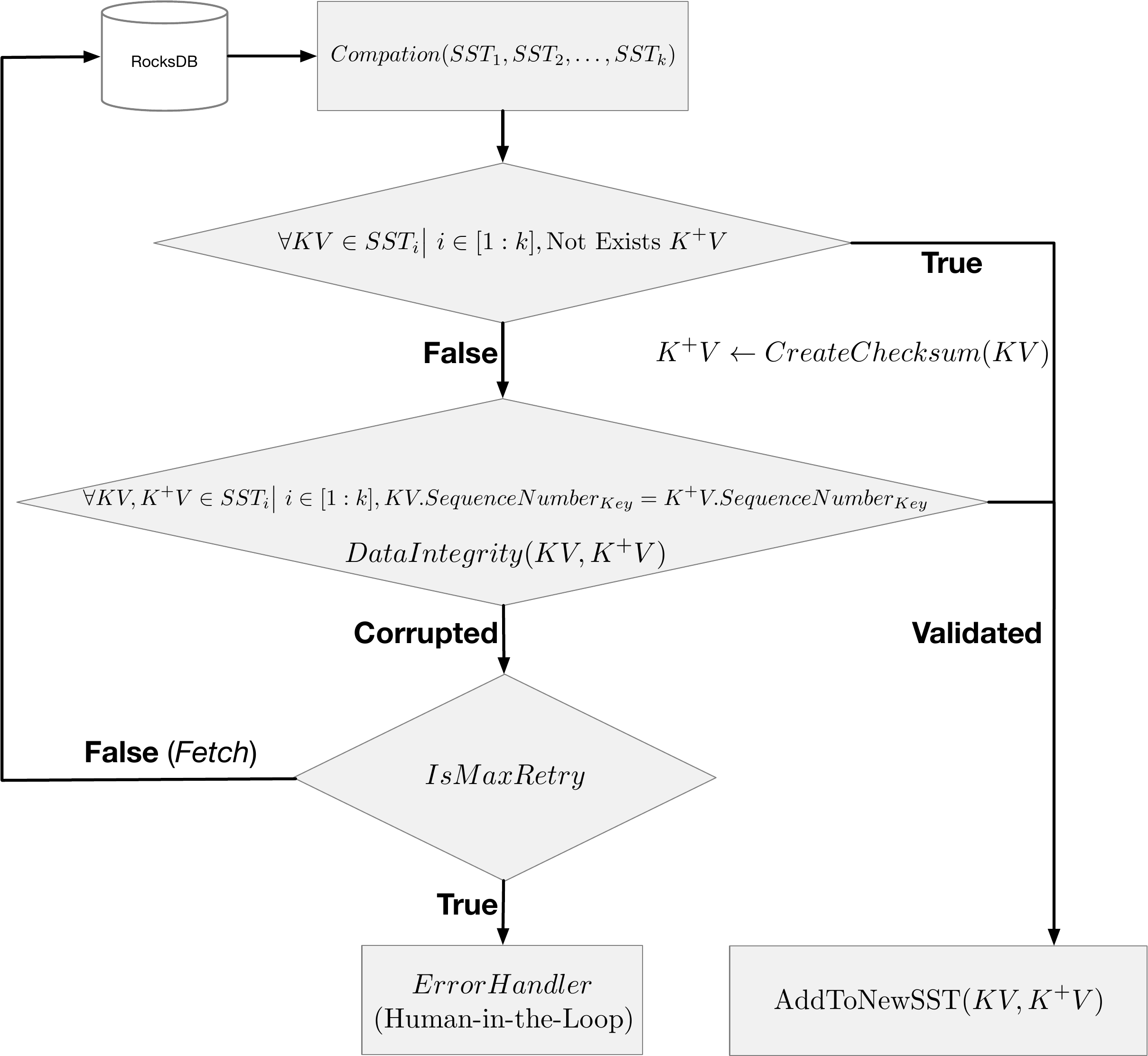}}
  \caption{The MetaHive design cycle within $Compation(.)$ and the repair process.}
  \label{fig:repair}
  \vspace{-1em}
\end{figure}

Next, we initiate the verification process for the key and its associated checksum, as depicted in Figure~\ref{fig:compaction}. Initially, if the checksum of the current $KV$, $CurrentChecksum$, matches the checksum stored in $K^+V$, the process returns $Validated$ to indicate the correctness of $KV$. If not, it checks the data in $K^+V$ by computing the checksum of its checksum and comparing it with the corresponding part. A match implies that $KV$ is erroneous, which prompts the module to return $Corrupted$. If there is no match, it recalculates the checksum of $CurrentChecksum$ and compares it with the checksum of the checksum in $K^+V$. If these are equal, it indicates that $K^+V$ is incorrect; Hence, it prevents the need to retrieve $KV$, which is generally quite large, and $K^+V$ can be recalculated using $CreateChecksum(.)$, returning a verdict $Validated$ for $KV$. Finally, if all comparisons fail, it indicates that both $K^+V$ and $KV$ are erroneous, and the process returns $Corrupted$.

\subsubsection{Repair Module Strategy}
This module uses the error detection module as a subprocess to identify erroneous data during the compaction process. As shown in Figure \ref{fig:repair}, for each key-value $KV$ in the SSTs that undergo compaction, if the associated $K^+V$ is located, it is forwarded to the error detection module (Figure \ref{fig:compaction}); otherwise, it is added to the new SST. If the data are validated, no further action is required and $KV$ and $K^+V$ are added to the new SST. Otherwise, a maximum retry counter is set, and the system retrieves the data specified by the error detection module. If the error persists after retries, the $ErrorHandler$ module is activated, responsible for deeper investigation. This module often involves human-in-the-loop, bringing in domain experts for further analysis. Given the heterogeneity of the cluster of nodes, there is no requirement to locate $K^+V$. If $K^+V$ is not found, the $CreateChecksum$ function is applied to $KV$, and both are incorporated into the SST. 

% \subsubsection{Extending MetaHive for any Metadata}
% The design we propose in MetaHive can be employed for any kind of per KV metadata. \textit{CreateChecksum} function can be changed to any function that one needs to create a metadata payload and modify the checksum part (\texttt{C} and \texttt{CC} in Figure~\ref{fig:compaction}) of the metadata payload. Other parts of the metadata creation remain unchanged to create metadata for heterogeneous key-value clusters which is cache-optimized in retrieval.  

\begin{table*}
\caption{Comparison of performance results of RocksDB with and without MetaHive}
\label{tab:perf}
\begin{tabular}{c|ccc|ccc}
\hline
\multirow{2}{*}{\textbf{Workload}} &
  \multicolumn{3}{c|}{\textbf{\begin{tabular}[c]{@{}c@{}}PUT Operation\\ (Nanosecond)\end{tabular}}} &
  \multicolumn{3}{c}{\textbf{\begin{tabular}[c]{@{}c@{}}GET Operation\\ (Nanosecond)\end{tabular}}} \\ \cline{2-7} 
 &
  \multicolumn{1}{c|}{\textbf{RocksDB}} &
  \multicolumn{1}{c|}{\textbf{MetaHive}} &
  \textbf{Difference (\%)} &
  \multicolumn{1}{c|}{\textbf{RocksDB}} &
  \multicolumn{1}{c|}{\textbf{MetaHive}} &
  \textbf{Difference (\%)} \\ \hline
\textbf{\begin{tabular}[c]{@{}c@{}}Read-Only\\ (0\% write rate)\end{tabular}} &
  \multicolumn{1}{c|}{\textbf{-}} &
  \multicolumn{1}{c|}{-} &
  - &
  \multicolumn{1}{c|}{172124} &
  \multicolumn{1}{c|}{172829} &
  \textbf{0.41\%} \\ \hline
\textbf{\begin{tabular}[c]{@{}c@{}}Read-Heavy\\ (5\% write rate)\end{tabular}} &
  \multicolumn{1}{c|}{83530} &
  \multicolumn{1}{c|}{84001} &
  \textbf{0.56\%} &
  \multicolumn{1}{c|}{163417} &
  \multicolumn{1}{c|}{164021} &
  \textbf{0.37\%} \\ \hline
\textbf{\begin{tabular}[c]{@{}c@{}}Update-Heavy\\ (50\% write rate)\end{tabular}} &
  \multicolumn{1}{c|}{100184} &
  \multicolumn{1}{c|}{100931} &
  \textbf{0.74\%} &
  \multicolumn{1}{c|}{159856} &
  \multicolumn{1}{c|}{160543} &
  \textbf{0.43\%} \\ \hline
\end{tabular}
\end{table*}

\section{Evaluations}
\label{sec:eval}
In this section, we first explain our experimental settings to evaluate MetaHive for metadata management. Then, we demonstrate its effectiveness in meeting the three metadata management objectives (defined in Section~\ref{sec:intro}) for a heterogeneous cloud cluster.

\subsection{Experimental Settings}
\subsubsection{Environment}
We implement MetaHive on the RocksDB version 8.1.1. We perform our evaluation on an Ubuntu 20.04 Linux machine with AMD Ryzen ThreadRipper Pro 5995WX 64-core 2.7GHz CPU and 256GB DDR4 RAM.

\subsubsection{Dataset and Workloads} We utilize YCSB~\cite{ycsb} to produce key-value pairs and simulate workloads. The key and value sizes are configured to 20 and 100 bytes, respectively. Our evaluations are conducted on the following YCSB workloads using a zipfian distribution: \textit{(1) Read-Only} with no PUT operations, \textit{(2) Read-Heavy} with 5\% PUT operations, and \textit{(3) Update-Heavy} with 50\% PUT operations. Each workload is executed for 2 million operations.
\subsection{Runtime Analysis} 
\subsubsection{Performance Overhead} 
Table~\ref{tab:perf} shows the performance results of different workload executions on RocksDB with and without MetaHive. The MetaHive version includes the insertion of checksum metadata and the execution of the error detection algorithm. We excluded 1\% of outlier data and utilized the median times of PUT and GET operations to measure performance.

Our results show that MetaHive has a negligible impact on GET operations with less than 0.5\% overhead on the latency. This is because MetaHive does not generate $K^+V$ (Section~\ref{sec:exs-payload}), therefore, fetching a key-value does not have the overhead of getting the metadata payload. We ran the same experiments by adding metadata to the value payload. In this case, we observe a slower throughput of more than 10\% through all three workloads.

MetaHive also has less than 0.8\% impact on PUT operations. We migrate the addition of checksum metadata to when the SST is created from the immutable MemTables (Figure~\ref{fig:RocksDB-arch}). This ensures that MetaHive does not need to locate the correct position of the keys for any new entries in the current MemTable. Writing the 	\textit{Immutable Memtable} to new SST files occurs as a background process, which minimally affects users' PUT operations. Also, as the keys are flushed in a sorted manner and the checksum metadata is always placed after the keys, there is no overhead for looking up the metadata place. 

\subsubsection{Memory and Storage Overhead}
MetaHive storage overhead is as follows: \textit{(1)} the key length plus one special character to put the metadata right after the key in the SST files, and \textit{(2)} eight bytes for the sequence number to find the corresponding entry in the metadata in clusters. The rest are the payloads that we need to put in the metadata. In our data integrity scenario, we add eight bytes of checksum and eight bytes of payload checksum to the metadata payload to check the data correctness.

MetaHive incurs minimal memory overhead during the compaction process, which is influenced by the number of keys within a cluster. By precalculating the checksum of each key and storing it in the map using its sequence number as the map's key (Section\ref{sec:alg}), the memory overhead is 16 Bytes per key in the cluster ($16 \times \vert C_{key}\vert$). Consequently, even with a cluster containing 50K keys, the overhead still remains below 0.001\%.

\subsection{Error Detection in Heterogeneous KV Cluster}
To evaluate the heterogeneity feature of MetaHive, we design an experiment with three RocksDB key-value nodes and one load-balancing node that distributes the data between RocksDB shards based on key prefixes. We added the MetaHive feature to one of the nodes, while two other KV nodes have the lower software version without MetaHive. We also embed a fault injection code into the MetaHive that modifies one bit of the checksum payload with a chance of 1\%. We also periodically assigned different prefixes to each shard to migrate the data between shards.

This experiment shows us that data from the old nodes got metadata when they migrated to the MetaHive node, and no metadata was transferred to the old nodes. Even when migrating a full SST from MetaHive to the old nodes, metadata was removed automatically without users' notice since they have Tombstone type. In addition, we observe faulty data detected before sending it to the new node for the first time it is undergoing compaction.

\section{Conclusion}
\label{sec:conclusion}
We introduce MetaHive, a powerful method to manage and access metadata in key-value stores, considering heterogeneity and privacy. MetaHive is utilized to incorporate checksum metadata into key-value entries in RocksDB. Our findings indicate that MetaHive identifies corrupted data with negligible performance impact in a heterogeneous key-value store cluster.

\bibliographystyle{ACM-Reference-Format}
\bibliography{main}

\end{document}